\documentclass[prb,preprint,12pt,superscriptaddress,floatfix,bibnotes,nofootinbib,unsortedaddress,preprintnumbers,amsmath,amssymb]{revtex4}

\usepackage{graphicx}% Include figure files
\usepackage{dcolumn}% Align table columns on decimal point
\usepackage{epstopdf}
\usepackage{xcolor} %\fcolorbox{yellow}{yellow}{ }

\usepackage{color}

\newcommand{\cm}{cm$^{-1}$}

\newcommand{\p}{^\prime}
\newcommand{\pp}{^{\prime\prime}}

\newcommand{\ai}{\textit{ab initio}}

\newcommand{\TROVE}{{\sc TROVE}}
\newcommand{\ohhh}{H$_3$O$^{+}$}
\newcommand{\oddd}{D$_3$O$^{+}$}
\newcommand{\ohhd}{H$_2$DO$^{+}$}
\newcommand{\ohdd}{HD$_2$O$^{+}$}

\newcommand{\Cv}[1]{${\mathcal C}_{#1{\rm v}}$}
\newcommand{\Dh}[1]{${\mathcal D}_{#1{\rm h}}$}

\newcommand{\3}{$_{3}$}

\def\deg{$^{\circ}$}

\begin{document}

\title{Radiative cooling of H$_3$O$^{+}$ and its deuterated isotopologues}

%\title{Stability of ro-vibrational states of H$_3$O$^{+}$, HD$_2$O$^{+}$, H$_2$DO$^{+}$ and D$_3$O$^{+}$}

\author{Vladlen V. Melnikov}%
\thanks{Corresponding author.
E-mail: melnikov@phys.tsu.ru} \affiliation{Siberian Institute of Physics \& Technology,
Tomsk State University, Tomsk, 634050 Russia}%
%Authors' institution and/or address\\
%This line break forced with \textbackslash\textbackslash

\author{Sergei N. Yurchenko}%
\affiliation{Department of Physics \& Astronomy, University College London, London WC1E~6BT, UK}

\author{Jonathan Tennyson}
\affiliation{Department of Physics \& Astronomy, University College London, London WC1E~6BT, UK}

\author{Per Jensen}
\affiliation{Physikalische und Theoretische Chemie,
Fakult\"at f\"ur Mathematik und Naturwissenschaften,
Bergische Universit{\"a}t Wuppertal, 42097 Wuppertal, Germany}

\date{\today}% It is always \today, today,
             %  but any date may be explicitly specified

\begin{abstract}

In conjunction with \textit{ab initio} potential energy and dipole moment surfaces for the electronic ground state, we have made
    a theoretical study of the radiative lifetimes for the hydronium ion H$_3$O$^{+}$  and its deuterated isotopologues. We compute the ro-vibrational energy levels and their associated wavefunctions
together with Einstein coefficients for the electric dipole transitions. A detailed analysis of the stability of the ro-vibrational states have been carried out
and the longest-living states of the hydronium  ions have been identified. We report
estimated radiative lifetimes and cooling functions for temperatures $<$~200~K.
A number of long-living meta-stable states are identified, capable of population trapping.

\end{abstract}

\maketitle

%Please use \dag to cite the ESI in the main text of the article.
%If you article does not have ESI please remove the the \dag symbol from the title and the footnotetext below.
%\footnotetext{$^{\ast}$~Corresponding author; E-mail: s.yurchenko@ucl.ac.uk }
%\footnotetext{\dag~Electronic Supplementary Information (ESI) available. See DOI: 10.1039/b000000x/}
%additional addresses can be cited as above using the lower-case letters, c, d, e... If all authors are from the same address, no letter is required

%\footnotetext{\ddag~Additional footnotes to the title and authors can be included %\textit{e.g.}\ `Present address:' or `These authors contributed equally to this work' as above %using the symbols: \ddag, \textsection, and \P. Please place the appropriate symbol next to %the author's name and include a \texttt{\textbackslash footnotetext} entry in the the correct %place in the list.}

%=============================================================================

%-----------------------------------------------------------------------------
\section{Introduction}

In the Universe, molecules  are found in a wide variety of environments: From diffuse
interstellar
clouds at very low temperatures to the atmospheres of planets, brown dwarfs and
cool stars which are significantly hotter.
In order to describe the evolution of the diverse,
complex environments, it is essential to have realistic predictions of the
radiative and cooling properties of the constituent molecules. Such predictions,
in turn, require
reasonable models for the energetics
of each molecular species present.

Although interstellar molecular clouds are usually characterised as cold, they
are mostly not fully thermalized. Whether a species attains thermal equilibrium
with the environment depends on the radiative lifetimes of its states and the
rate of collisional excitations to the states: This is normally
characterised by the critical density. In non-thermalized regions, radiative
lifetimes are also important for modelling the maser activity observed for
many species.
 The long lifetimes associated with certain excited states can lead to
population trapping and non-thermal, inverted distributions. Such
unexpected state distributions have been observed
for the H$^+_3$ molecule both in space\cite{02GoMcGe.H3+, 05OkGeGo.H3+} and in
the laboratory.\cite{02KrKrLa.H3+,04KrScTe.H3+}

Dissociative recombination of hydronium
     H$_3$O$^{+}$
has been extensively studied in ion storage rings.%
\cite{96AnHeKe.H3O+,00NeKhRo.H3O+,00JeBiSa.H3O+,10BuStMe.H3O+,10NoBuSt.H3O+}
The lifetimes calculated in the present work suggest
 that H$_3$O$^{+}$ and it isotopologues
will exhibit
 population trapping in a manner similar to that observed
for H$_3^+$ in storage rings. Dissociative recombination of hydronium
has been postulated as a possible cause of emissions from super-excited
water in cometary comae\cite{09BaMiDe}
%\cite{jt452}
and as the mechanism for a spontaneous
infrared water laser.\cite{01SaKeWa.H3O+}

Hydronium and its isotopologues play an important role in planetary and
interstellar chemistry.\cite{00JeBiSa.H3O+,01GoCexx.H3O+} These molecular ions
are found to exist abundantly in both diffuse and dense molecular clouds as well
as in comae. Moreover, H$_3$O$^{+}$ is a water indicator and can be used to
estimate water abundances when the direct detection is
unfeasible.\cite{92PhVaKe.H3O+} Consequently,
 the ions have been the
subject of numerous theoretical and experimental studies (see, for example,
Refs.~%
\onlinecite{00JeBiSa.H3O+,01GoCexx.H3O+, 92PhVaKe.H3O+,  73LiDyxx.H3O+,
80FeHaxx.H3O+, 82SpBuxx.H3O+, 83BeGuPf.H3O+, 83BoRoRe.H3O+, 85BoDeDe.H3O+, 84DaHaJo.H3O+, 84BuAmSp.H3O+,
85BeSaxx.H3O+, 85LiOkxx.H3O+, 85PlHeEr.H3Op, 86DaJoHa.H3O+, 86LiOkSe.H3O+, 87GrPoSa.H3O+, 88HaLiOk.H3O+,
88VeTeMe.H3O+, 90OkYeMy.H3O+, 90PeNeOw.H3O+,
91HoPuOk.H3O+, 97UyWhOk.H3O+, 99ArOzSa.H3O+, 00ChJuGe.H3O+, 01AiOhIn.H3O+,
02ErSoDo.H3O+, 02CaWaZu.H3O+, 03RaMiHa.NH3, 03HuCaBo.H3O+, 05YuBuJe.H3O+,
06DoNexx.H3O+, 08FuFaxx.H3O+, 09YuDrPe.H3O+, 10MuDoNe.H3O+,
12PeWeBe.H3O+,10BuStMe.H3O+} and references therein) mainly devoted to the
spectroscopy and chemistry of the species.

Whereas the cooling function of the H$_3^{+}$ ion
has been extensively studied by Miller et al,\cite{96NeMiTe.H3+,10MiStMe.H3+,13MiStTe.H3+}
no information about the
radiative and cooling properties of  H$_3$O$^{+}$ and its deuterated isotopologues
has been available thus far.
 In the present
work, we remedy this situation by determining theoretically
 the ro-vibrational states of the ions
 H$_3$O$^{+}$,
 H$_2$DO$^{+}$,
 HD$_2$O$^{+}$, and
 D$_3$O$^{+}$.
  We use    \ai\  potential energy (PES) and dipole moment surfaces
(DMS) for the ground electronic states of H$_3$O$^{+}$ from
Ref.~\onlinecite{15OwYuPo.H3O+} to compute
for each of the four ions considered here,
ro-vibrational energy levels, the accompanying wavefunctions, and Einstein coefficients
for the relevant ro-vibrational (electric dipole) transitions
by means of
the nuclear-motion program TROVE.\cite{TROVE}
Lifetimes of individual ro-vibrational states are calculated and analyzed together with
the overall cooling rates.
Recently, the same methodology was used to estimate the sensitivities of
hydronium-ion  transition frequencies to a possible time variation of the proton-to-electron mass
ratio.\cite{15OwYuPo.H3O+}

We present a detailed analysis of the stability of the ro-vibrational states of the
hydronium ions and
identify the states with the longest lifetimes.  This study is based on the
methodology\cite{16TeHuNa.method} developed very recently as part of the ExoMol
project. \cite{12TeYuxx.db} The ExoMol project aims at a comprehensive
description of spectroscopic properties of molecules important for atmospheres
of exoplanets and cool stars. The molecular lifetimes and cooling functions
determined for
 H$_3$O$^{+}$ and its deuterated isotopologues
 in the present work are
available in
the new ExoMol data format.\cite{16TeYuAl.db}

%-----------------------------------------------------------------------------
\section{Theory and computation}

The ro-vibrational energy levels and wavefunctions of the ions under study were
calculated variationally with      the \TROVE\ program%
\cite{TROVE,15YaYuxx.method} in a manner similar to the successful calculations
previously carried out for several other
 XY\3\ pyramidal molecules%
\cite{10YuCaYa.SbH3,13SoYuTe.PH3,13UnTeYu.SO3,14UnYuTe.SO3,
15SoAlTe.PH3,15AdYaYuJe.CH3} including ammonia NH$_3$ (see Refs.~\onlinecite{09YuBaYa.NH3,11YuBaTe.NH3}),
which exhibits the same large-amplitude, `umbrella-flipping' inversion motion
  as \ohhh.
 The
inversion barrier of \ohhh\ is 650.9~\cm,\citep{04RaNoVa.H3O+} which is lower than the 1791~\cm\ found
for NH$_3$.\cite{03RaMiHa.NH3} As a result, the \ohhh\ inversion splitting of the
ro-vibrational ground state,
 55.35~\cm\ (Ref.~\onlinecite{99TaOkxx.H3O+}), is significantly larger than the
0.793~\cm\ (Ref.~\onlinecite{03RaMiHa.NH3}) of        NH$_3$.

We have used \ai\ PES and DMS of \ohhh\ (Refs.~\onlinecite{15OwYuPo.H3O+,15Yuxxxx.15NH3}),
computed at    the MRCI/aug-cc-pwCV5Z (5Z) and MRCI/aug-cc-pwCVQZ (QZ) levels of
theory, respectively. Complete basis set (CBS) extrapolation was used to obtain the
Born-Oppenheimer PES (see Ref.~\onlinecite{15OwYuPo.H3O+} for details).

The basis set used in the variational computations
of the ro-vibrational states is defined
in Ref.~\onlinecite{TROVE}.
In all calculations of the present work, the orders of kinetic and
potential energy expansions were set to 6 and 8, respectively. We used
Morse-type basis functions for the stretching modes and numerical basis functions
(numerical solutions of corresponding 1D problem obtained within the framework
of the Numerov-Cooley scheme) for the bending vibrations. The \ai\ equilibrium structure
of \ohhh\ is characterized by an  O-H bond length of  0.9758~{\AA}
and H-O-H angle of       111.95\deg. The vibrational basis set is controlled by
the polyad number defined by
\begin{equation}
\label{e:polyad}
P = 2(v_1 + v_2 + v_3) + v_4 + v_5 + v_6/2,
\end{equation}
where $v_1$, $v_2$, $v_3$ represent the quanta of the stretching motion, $v_4$ and
$v_5$ are those of the asymmetric bending motion and $v_6$ is that of the primitive
basis set functions for the inversion. In the present work,
 the maximum polyad  number $P_{\rm max}$
 (where $P$ $\leq$ $P_{\rm max}$) was
set to 28. The ro-vibrational basis sets used for \ohhh\ and \oddd\
were symmetrized according with the \Dh{3}(M) molecular symmetry group \cite{98BuJexx.method}, while the
calculations for the asymmetric isotopologues \ohhd\ and \ohdd\ were done under the \Cv{2}(M) symmetry.
The computational details of the basis set construction can be found
in Ref.~\onlinecite{09YuBaYa.NH3} as well as the details of the calculations of the
Einstein coefficients $A_{if}$.
The latter were computed for all possible
initial, $i$, and final, $f$, states lying less than 600~cm$^{-1}$ above
the zero point energy  with $J\le  7$. According to our estimations, this should be
sufficient to describe the populations of the ro-vibrational states at thermodynamic temperatures up to 200~K.
In this work we concentrate on the low energy applications. Higher temperatures would require higher rotational excitations $J$
and therefore more involved calculations.

The lifetimes of the states were computed as\cite{16TeHuNa.method}
\begin{equation}
\label{eq: tauval}
  \tau_i = \frac{1}{\sum_f A_{if}},
\end{equation}
where the summation is taken over all possible \textit{final} states $f$
with energies lower than that of
 the given \textit{initial} state $i$. The Einstein coefficients (in units of 1/s)
are defined as follows:\cite{92GaRoxx}
\begin{equation}
  A_{if} = \frac{64\pi^{4}\tilde{\nu}_{if}^{3}}{3h (2 J_f + 1) }\,  \sum_{m_i,m_f}
  \sum_{A=X,Y,Z} |\langle \Psi^{f} | \bar{\mu}_{A} | \Psi^{i} \rangle |^{2},
\label{e:A}
\end{equation}
%
%
%\begin{equation*}
%  A_{if} = \frac{64\pi^{4}\tilde{\nu}_{if}^{3}}{3h}\,  \sum_{m_f}
%  |\langle \Psi^{f} | \bar{\bm{\mu}} | \Psi^{i} \rangle |^{2},
%\label{e:A}
%\end{equation*}
%
%
where  $h$ is Planck's constant, $\tilde{\nu}_{if}$ (\cm) is the wavenumber of
the line, (\(hc \, \tilde{\nu}_{if} = E_{f} -E_{i}\)), $J_{f}$ is the rotational quantum number for
the final state, \(\Psi^{i}\) and \(\Psi^{f}\)  represent the ro-vibrational
eigenfunctions of the final and initial states respectively, $m_i$ and $m_f$ are the corresponding projections of the total angular momentum on the $Z$ axis, and $\bar{\mu}_{A}$ is the electronically averaged component of the dipole moment (Debye) along the space-fixed axis \(A=X,Y,Z\) (see also Ref.~\onlinecite{05YuThCa.method}).

The cooling function $W(T)$ is the total power per unit solid angle emitted by a molecule at temperature
$T$;
it is given by the following expression:\citep{16TeHuNa.method}
\begin{equation}
\label{e:cooling}
  W(T) = \frac{1}{4\pi Q(T)} \sum_{i,f} A_{if}\, h c\, \tilde{\nu}_{if}\, (2 J_i+1)
\, g_i\,  \exp\left( -\frac{c_2 \tilde{E}_i}{T} \right) ,
\end{equation}
where
 $g_i$ is the
nuclear spin statistical weight factor of the state $i$.
 In the                   Boltzmann factor
 $\exp( -c_2 \tilde{E}_i/T )$,
$\tilde{E}_i$ ($=$ $E_i/hc$) is the term value
of state $i$ and $c_2 = h c/k$ is the second radiation constant ($k$ is the Boltzmann constant).
 The partition function $Q(T)$ is defined as
\begin{equation}
  Q(T) = \sum_{i} g_i  (2J_i+1) \exp\left( -\frac{c_2 \tilde{E}_i}{T} \right) .
\end{equation}
Partition functions were computed for each ion employing the      sets of
 ro-vibrational energies obtained with the chosen basis set.

The same PES and DMS were used for each isotopologue, which means that
no allowance was made for the breakdown of the Born-Oppenheimer approximation.
The energies of
 \ohhh\ and the three deuterated isopologues
 are very different not only due to the mass
changes, but also due to the different symmetries these species belong to
and the ensuing nuclear spin statistics. We
use the molecular symmetry group\cite{98BuJexx.method}
\Dh{3}(M) to classify the ro-vibrational
states of the highest-symmetry species \ohhh\ and \oddd, and \Cv{2}(M) for the
lower-symmetry
isotopologues \ohhd\ and \ohdd. The differences in the Einstein
coefficients are also quite substantial, especially between the
\Dh{3}(M) and the
 \Cv{2}(M)
 isotopologues. In general, isotope substitution in ions often leads to
large changes in ro-vibrational intensities. This is because these intensities
depend on the components of the electric dipole moment in the molecule-fixed axis
system which, by definition, has its origin in the nuclear center-of-mass. Upon
isotopic substitution, the
center-of-mass,
and thus the origin of the molecule-fixed axis system,
 are displaced. For a neutral molecule (i.e., a  molecule with no net charge)
this does not change the dipole moment components but, for an ion,
they do (see, for example, Ref.~\onlinecite{88Jensen.CH2}).
Owing to intensities and Einstein coefficients  changing much upon isotopic substitution,
the lifetimes of the ro-vibrational states are
 expected to vary strongly with isotopologue for \ohhh.
Selection rules for $J$ are
\begin{equation}
 J\p - J\pp = 0,\pm 1  \quad {\rm and} \quad  J\p + J\pp > 0.
\end{equation}
The symmetry selection rules for \ohhh\
and \oddd\ are
\begin{equation}
 A_1\p \leftrightarrow A_1\pp , \quad  A_2\p \leftrightarrow A_2\pp , \quad
  E\p \leftrightarrow E''
\end{equation}
(where, for \ohhh\, levels of
 $A_1\p$ and $A_1\pp$ symmetry
are missing\cite{98BuJexx.method} so that the associated selection rule
is irrelevant),
while those for \ohhd\ and \ohdd\ are
\begin{equation}
 A_1 \leftrightarrow A_2, \quad B_1 \leftrightarrow B_2.
\end{equation}

%n decreasing order of the symmetry degeneracy gns: ortho (E, gns = 16), meta (A1, gns = 10) and para (A2, gns = 1).

Nuclear spin statistics\cite{98BuJexx.method} result in three
 distinct spin species of \oddd, \textit{ortho}
(ro-vibrational symmetry
 $A_1\p$ or  $A_1\pp$, nuclear spin statistical weight factor
  $g_{\rm ns}=10$), \textit{meta} ($E\p$ or $E\pp$,
 $g_{\rm ns}=8$) and \textit{para}
 ($A_2\p$ or  $A_2\pp$,
 $g_{\rm ns}=1$).
As mentioned above,
 $A_1\p$ and $A_1\pp$ ro-vibrational states are missing for
 \ohhh\ and there are only \textit{ortho}
 ($A_2\p$ or  $A_2\pp$,
 $g_{\rm ns}=4$) and \textit{para}
   ($E\p$ or $E\pp$,
 $g_{\rm ns}=2$) states.
 The \Cv{2}(M)
 isotopologues
 \ohhd\ and \ohdd\ have
  \textit{ortho} and \textit{para} states: For \ohhd,
$B_1$ and $B_2$ states are \textit{ortho} with $g_{\rm ns}=9$, and $A_1$ and $A_2$ states are \textit{para} with
$g_{\rm ns}=3$. For \ohdd, the \textit{ortho}-\textit{para} states are interchanged relative to \ohhd:
$A_1$ and $A_2$ states are now \textit{ortho} with $g_{\rm ns}=12$ whereas $B_1$ and $B_2$ states are \textit{para} with
$g_{\rm ns}=6$.

%Since Eqs.~(\ref{eq:A21}--\ref{eq:partition}) depend on energy eigenvalues, corresponding wavefunctions, electric dipole moment %and symmetries along with nuclear spin statistics. All these properties are affected when constituent atoms are substituted by %their isotopes.

%-----------------------------------------------------------------------------
\section{Results and discussion}

In order to validate our description of the energetics of \ohhh\ and its
deuterated isotopologues, we compare
in
Table~\ref{tab:Expt} calculated vibrational energies for these molecules with the available,
experimentally derived values. In view of the fact that the calculations are based
on a purely \ai\ PES, the agreement between theory and experiment is excellent.
The results suggest that the \ai\ data used in the present work, in conjunction
with the variational \TROVE\ solution of the ro-vibrational Schr\"odinger equation, are
more than adequate for obtaining accurate lifetimes of the molecules studied.

%especially considering that matrix elements of the dipole moment are not very sensitive to the quality of the underlying PES.

\begin{table*}[!htbp]
\small
  \tabcolsep=0.5cm
  \caption{\label{tab:Expt} Available, experimentally derived vibrational
term values of \ohhh\ and its deuterated isotopologues
(in cm$^{-1}$) compared to theoretical values from the present work}
  \renewcommand{\arraystretch}{0.75}

  \begin{tabular*}{\textwidth}{@{\extracolsep{\fill}}lcrcrr}

  \hline\hline
  State & Sym. & \multicolumn{1}{c}{Exp.} & \multicolumn{1}{c}{Ref.} & \multicolumn{1}{c}{Calc.} & \multicolumn{1}{c}{Exp.$-$Calc.} \\ \hline

  \multicolumn{6}{l}{H$_3$O$^+$} \\

  $\nu_2^+$   & $A_1$ &  581.17& [\onlinecite{86LiOkSe.H3O+}] &  579.07 &  2.10 \\
  $2\nu_2^+$  & $A_1$ & 1475.84& [\onlinecite{86DaJoHa.H3O+}] & 1470.67 &  5.17 \\
  $\nu_1^+$   & $A_1$ & 3445.01& [\onlinecite{99TaOkxx.H3O+}] & 3442.61 &  2.40 \\
  $\nu_3^+$   &  $E$  & 3536.04& [\onlinecite{99TaOkxx.H3O+}] & 3532.58 &  3.46 \\
  $\nu_4^+$   &  $E$  & 1625.95& [\onlinecite{87GrPoSa.H3O+}] & 1623.02 &  2.93 \\
  $0^-$       & $A_1$ &   55.35& [\onlinecite{99TaOkxx.H3O+}] &   55.03 &  0.32 \\
  $\nu_2^-$   & $A_1$ &  954.40& [\onlinecite{86LiOkSe.H3O+}] &  950.94 &  3.46 \\
  $\nu_1^-$   & $A_1$ & 3491.17& [\onlinecite{99TaOkxx.H3O+}] & 3488.32 &  2.85 \\
  $\nu_3^-$   &  $E$  & 3574.29& [\onlinecite{99TaOkxx.H3O+}] & 3571.04 &  3.25 \\
  $\nu_4^-$   &  $E$  & 1693.87& [\onlinecite{87GrPoSa.H3O+}] & 1690.65 &  3.22 \\

  \multicolumn{6}{l}{D$_3$O$^+$} \\

  $\nu_2^+$   & $A_1$ &  453.74& [\onlinecite{90PeNeOw.H3O+}]&   451.58 &  2.16 \\
  $\nu_3^+$   & $ E $ & 2629.65& [\onlinecite{90PeNeOw.H3O+}]&  2627.14 &  2.51 \\
  $0^-$       & $A_1$ &   15.36& [\onlinecite{98ArOzSa.H3O+}]&    15.38 & -0.02 \\
  $\nu_2^-$   & $A_1$ &  645.13& [\onlinecite{90PeNeOw.H3O+}]&   642.79 &  2.34 \\
  $\nu_3^-$   & $ E $ & 2639.59& [\onlinecite{90PeNeOw.H3O+}]&  2637.10 &  2.49 \\

  \multicolumn{6}{l}{H$_2$DO$^+$} \\

  $0^-$       & $B_1$ &     40.56& [\onlinecite{08FuFaxx.H3O+}]&      40.39 &   0.17 \\
  $\nu_1^+$   & $A_1$ &   3475.97& [\onlinecite{06DoNexx.H3O+}]&    3473.27 &   2.70 \\
  $\nu_1^-$   & $B_1$ &   3508.63& [\onlinecite{06DoNexx.H3O+}]&    3505.51 &   3.12 \\
  $\nu_3^+$   & $B_2$ &   3531.50& [\onlinecite{06DoNexx.H3O+}]&    3528.07 &   3.43 \\
  $\nu_3^-$   & $A_2$ &   3556.94& [\onlinecite{06DoNexx.H3O+}]&    3553.63 &   3.31 \\

  \multicolumn{6}{l}{HD$_2$O$^+$} \\

  $0^-$       & $B_1$ &   26.98& [\onlinecite{08FuFaxx.H3O+}]&    26.92 &  0.06 \\

  \hline\hline

  \end{tabular*}
%  \parbox{10.7cm}
  %{\scriptsize
  %\begin{flushleft}
  %$^{\rm a}$ Ref.~\onlinecite{86LiOkSe.H3O+}. \\
  %$^{\rm b}$ Ref.~\onlinecite{86DaJoHa.H3O+}. \\
  %$^{\rm c}$ Ref.~\onlinecite{99TaOkxx.H3O+}. \\
  %$^{\rm d}$ Ref.~\onlinecite{87GrPoSa.H3O+}. \\
  %$^{\rm e}$ Ref.~\onlinecite{90PeNeOw.H3O+}. \\
  %$^{\rm f}$ Ref.~\onlinecite{98ArOzSa.H3O+}.
  %\end{flushleft}}

\end{table*}

\begin{figure*}[!htbp]

\begin{tabular}{lr}
  \includegraphics[width=8cm]{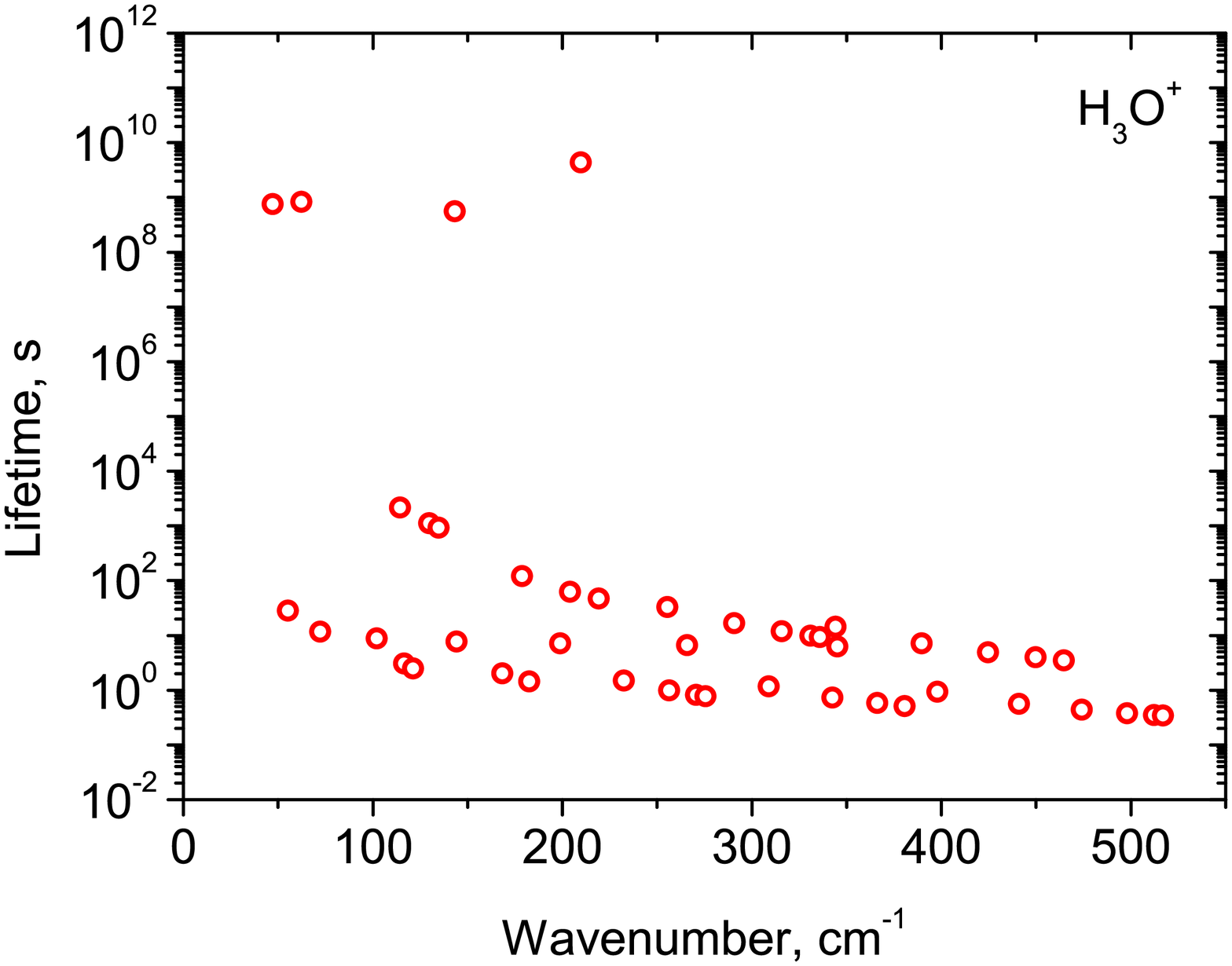} &
  \includegraphics[width=8cm]{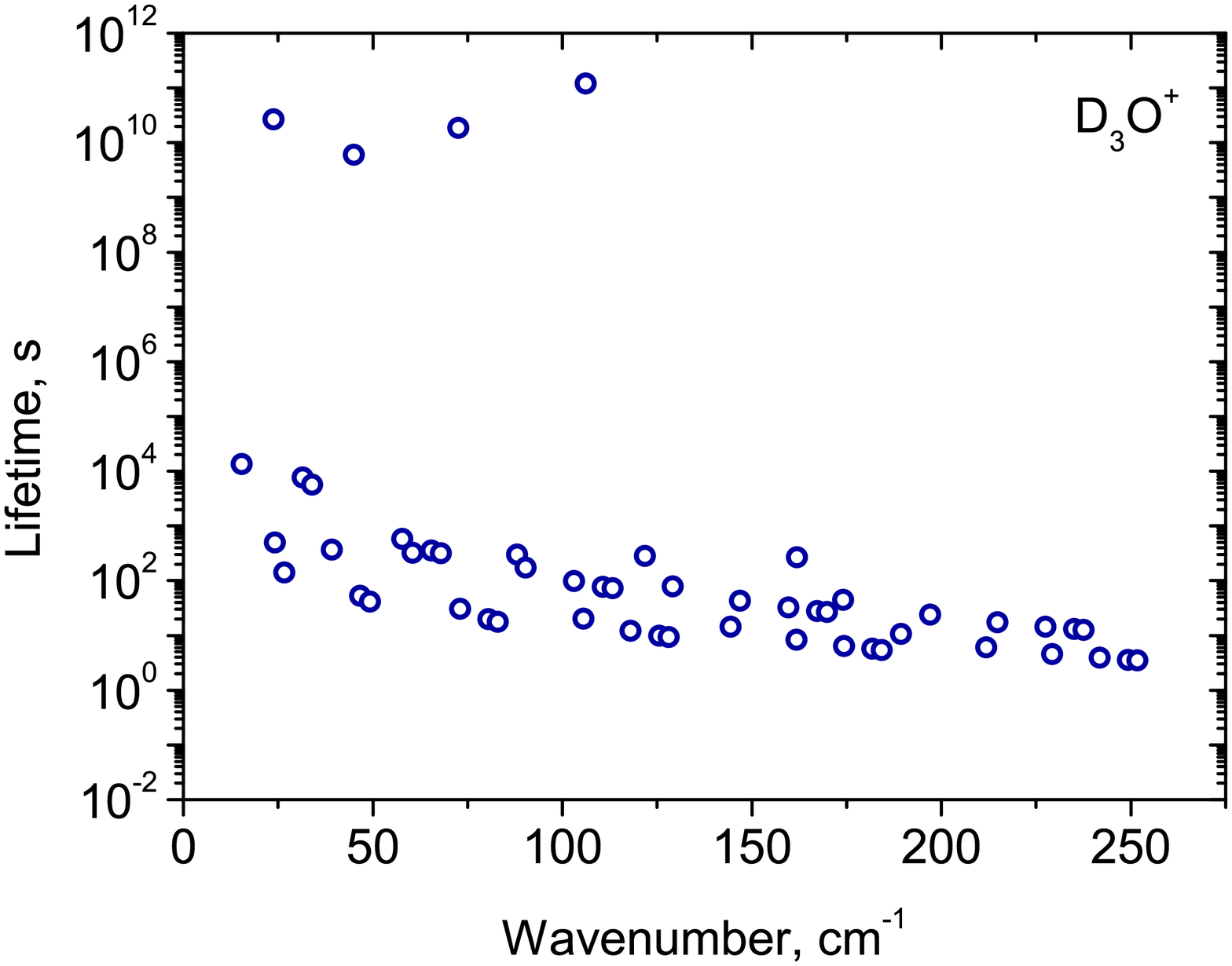} \\
  \includegraphics[width=8.26cm]{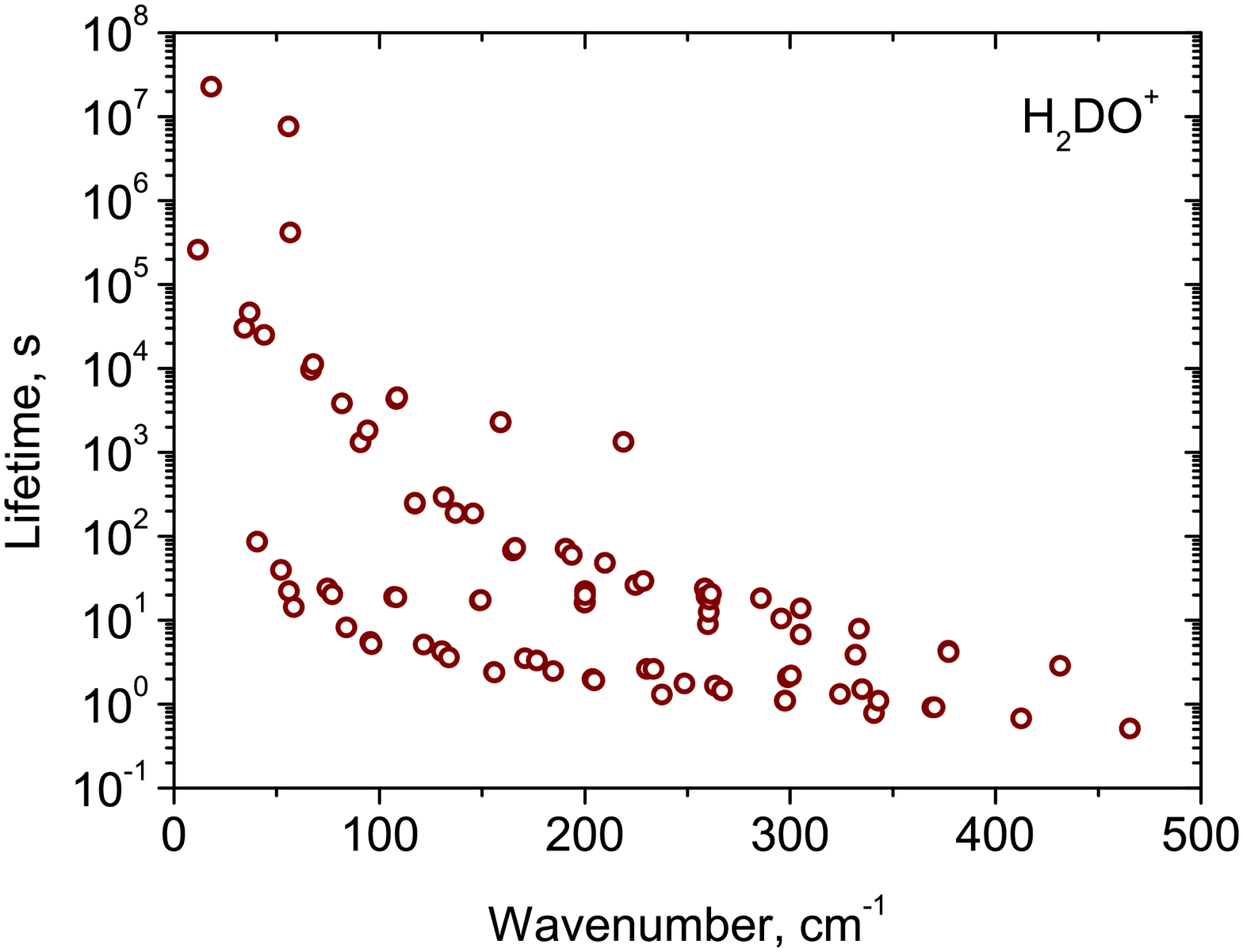} &
  \includegraphics[width=8cm]{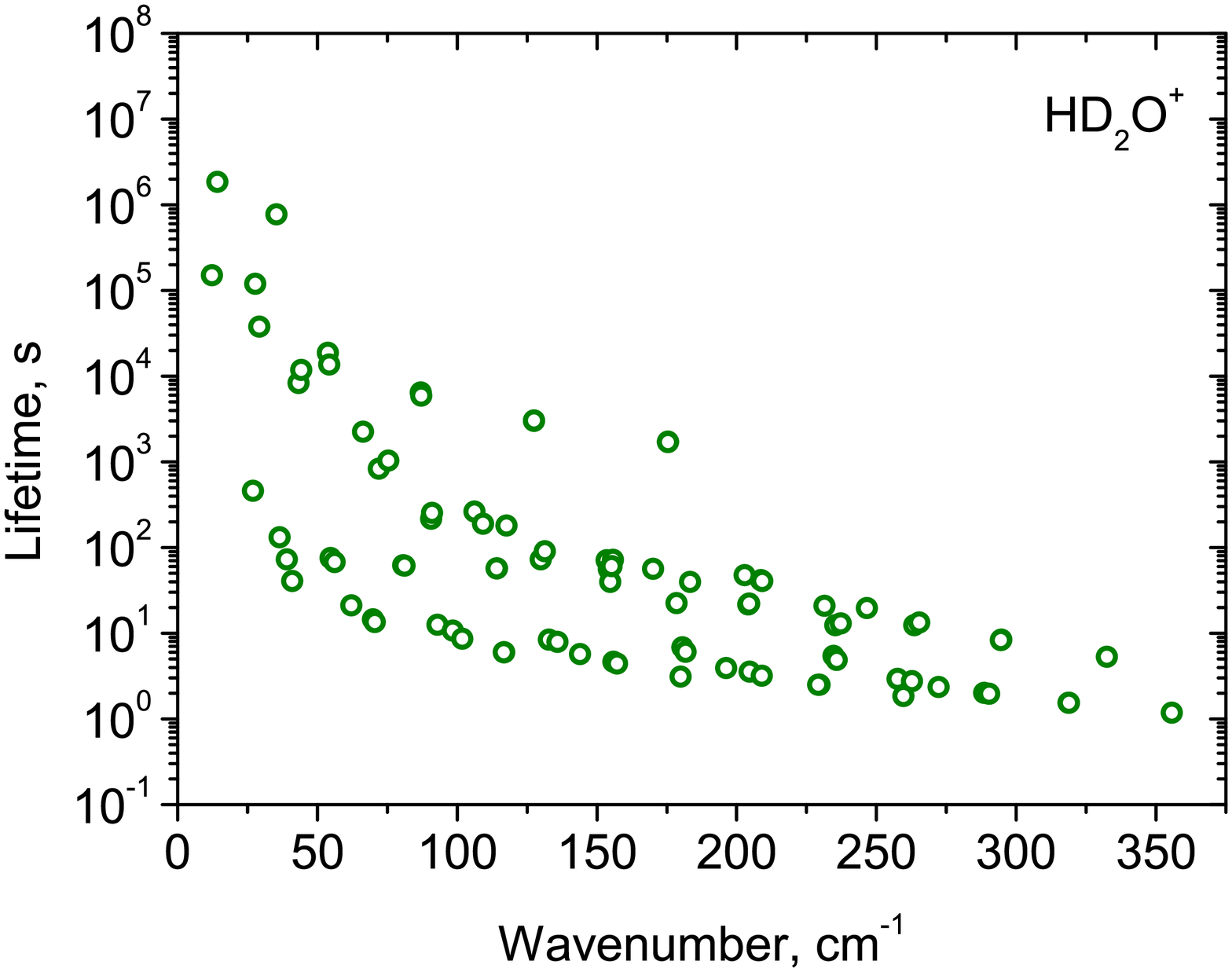}\\
\end{tabular}

%\begin{tabular}{cc}
%  \includegraphics[width=8cm]{h3op-1c-j7.eps} & \includegraphics[width=8cm]{d3op-1c-j7.eps} \\
%
%  \includegraphics[width=8cm]{h2dop-1c-j7.eps}& \includegraphics[width=8cm]{hd2op-1c-j7.eps}\\
%
%\end{tabular}

  \caption{\label{fig:lifetime-1} Calculated
lifetimes $\tau$ of the ro-vibrational states ($J\le 7$) of H$_3$O$^+$ and its deuterated
isotopologues. The lifetime values are plotted  logarithmically.}

\end{figure*}

In Fig.~\ref{fig:lifetime-1}, we plot the lifetimes calculated for the
ro-vibrational states of
\ohhh\ and its deuterated isotopologues against the associated term values $\tilde{E}$
($J \le 7$, $\tilde{E}$ $<$ 600~cm$^{-1}$).
In general,
the lifetimes exhibit the expected gradual decrease  with increasing term value.
 The complete list of lifetimes for all
four isotopologues are given as supplementary material to this paper.

%\red{CAN WE PUT THE FOLLLOWING INTO A TABLE?}
%The most long-living states identified are the following.
%For H$_3$O$^+$ there are
%$(0,0,A_2'')$, $\tau=23.9$ years;
%$(1,1,E')$, $\tau=26.2$ years;
%$(3,3,A_1')$, $\tau=17.7$ years;
%$(4,1,E'')$, $\tau=140$ years.
%For D$_3$O$^+$ there are
%$(1,1,E')$, $\tau=857$ years;
%$(2,1,E')$, $\tau=190.4$ years;
%$(3,1,E')$, $\tau=594.4$ years;
%$(4,1,E'')$, $\tau=3816$ years.
%For H$_2$DO$^+$ there are
%$(1,1,B_2)$, $\tau=3$ days;
%$(2,0,A_1)$, $\tau=265.4$ days;
%$(2,2,A_1)$, $\tau=89.1$ days;
%$(1,1,A_2)$, $\tau=4.8$ days.
%And for HD$_2$O$^+$ there are
%$(1,1,B_2)$, $\tau=1.8$ days;
%$(0,0,B_1)$, $\tau=21.6$ days;
%$(2,1,B_2)$, $\tau=1.4$ days;
%$(1,1,A_1)$, $\tau=8.9$ days.

\begin{table}[!htbp]
  \tabcolsep=0.5cm
  \caption{\label{tab:LF} Lifetimes $\tau$ for the longest-lived states of the
\ohhh, \oddd, \ohhd, and \ohdd\ ions. All states listed are rotational states in
the vibrational ground state (i.e., the inversion state $0^+$). The states are labeled
 $(J,K,\Gamma)$ and $(J_{K_a,K_c},\Gamma)$ for
\Dh{3}(M) and
 \Cv{2}(M) isopologues, respectively, with
$\Gamma$ as the symmetry of the state in the respective group.}
  \renewcommand{\arraystretch}{0.75}
  \begin{tabular}{lrr}
  \hline\hline
  State & \multicolumn{1}{c}{Term value, cm$^{-1}$} & \multicolumn{1}{c}{$\tau$} \\ \hline
\multicolumn{1}{l}{\ohhh}  & & \multicolumn{1}{c}{(years)}         \\
$(1,0,A_2\p)$         &$   22.47 $&$             \infty  $  \\
$(1,1,E\pp)$          &$   17.38 $&$             \infty  $  \\
$(3,3,A_2\pp)$        &$   88.96 $&$             \infty  $  \\
 $(5,5,E\pp)$         &$  209.58 $&$           140.1  $ \\
 $(2,1,E\pp)$         &$   62.29 $&$            26.2  $ \\
 $(2,2,E\p)$          &$   47.03 $&$            23.9  $ \\
 $(4,4,E\p)$          &$  143.15 $&$            17.7  $ \\
%H3O+
%E"        1    209.584753   ( E" ;  5  5  0) ( A1';   0   0   0   0   0   0)      0.99 (   0   0   0   0   0   0   0) (     1)    tau = 4.41647e+09 s = 140.1 years
\hline
\multicolumn{1}{l}{\oddd}  & & \multicolumn{1}{c}{(years)}        \\
$(0,0,A_1\p)$         &$  0.00  $&$             \infty  $  \\
$(1,0,A_2\p)$         &$  11.33 $&$             \infty  $   \\
$(1,1,E\pp)$          &$  8.78  $&$             \infty  $  \\
$(3,3,A_2\pp)$        &$  45.02 $&$             \infty  $  \\
$(5,5,E\pp)$          &$ 106.15 $&$         3816.0 $  \\
$(2,2,E\p)$           &$  23.79 $&$          857.1 $ \\
$(4,4,E\p)$           &$  72.48 $&$          594.4 $  \\
$(3,3,A_1\pp)$        &$  45.02 $&$          190.4 $  \\
%  E"        1    106.154185   ( E" ;  5  5  0) ( A1';   0   0   0   0   0   0)      0.99 (   0   0   0   0   0   0   0) (     1)    tau = 1.20341e+11 s = 3816.0 years
\hline
\multicolumn{1}{l}{\ohhd} & &  \multicolumn{1}{c}{(days)}         \\
$(0_{0,0},A_1)$           &$   0.00 $&$             \infty $   \\
$(1_{1,1},B_1)$           &$  15.70 $&$             \infty $   \\
$(1_{1,0},B_2)$           &$  18.07 $&$          265.4 $  \\
$(2_{2,1},A_2)$           &$  55.82 $&$           89.1 $  \\
$(2_{2,0},A_1)$           &$  56.60 $&$            4.8 $  \\
$(1_{0,1},A_2)$           &$  11.69 $&$            3  $ \\
\hline
\multicolumn{1}{l}{\ohdd} & &  \multicolumn{1}{c}{(days)}         \\
$(0_{0,0},A_1)$           &$  0.00  $&$             \infty  $  \\
$(1_{1,1},B_1)$           &$  9.53  $&$             \infty  $  \\
$(1_{1,0},B_2)$           &$  14.24 $&$           21.6  $ \\
$(2_{2,1},A_2)$           &$  35.35 $&$            8.9  $ \\
$(1_{0,1},A_2)$           &$  12.19 $&$            1.8  $ \\
$(2_{2,0},A_1)$           &$  27.77 $&$            1.4  $ \\
  \hline\hline
  \end{tabular}

\end{table}

Lifetimes $\tau$  of the longest-lived states of the ions are
compiled   in Table~\ref{tab:LF}.
The lowest-lying state of each of the spin species
\textit{ortho} and
\textit{para}
(and \textit{meta} for \oddd) has an infinitely long lifetime; it has no state to decay to.
% $(1_{1,1}^+,B_1)$ state
%for both \ohhh\ and \oddd\ cannot decay via any dipole-allowed transitions so shows
%up as having an infinite lifetime.

Low-lying, purely rotational states with low $J$ values have the longest lifetimes;
they  have the smallest numbers decay of channels
and/or the lowest probability for spontaneous emission transitions.
 The higher-symmetry species
\ohhh\ and \oddd\ (with \Dh{3}(M) symmetry) have more restrictive
selection rules than the
\Cv{2}(M) species
\ohdd\ and \ohhd, and so
\ohhh\ and \oddd\ states live
 significantly longer (typically tens to hundreds of years) compared
to the day-long lifetimes of \ohdd\ and \ohhd.
 Thus,        \oddd\ has three meta-stable states
with lifetimes longer than 100 years. The longest-lived of
these,  with $\tau$ $=$ 3816~years,  is the rotational state
$(J=5,K=5,E'')$ of the vibrational ground state. In comparison,
%these,  with $\tau$ $=$ 857~years,  is the rotational state
%$(J=2,K=2,E')$ of the vibrational ground state. In comparison,
the longest-lived meta-stable state
of \ohhd\, the rotational state
 $(J_{K_a,K_c},\Gamma)$ $=$
$(1_{1,0},B_2)$, has a lifetime of 265 days.

\begin{figure}%[!htbp]

  \includegraphics[width=9cm]{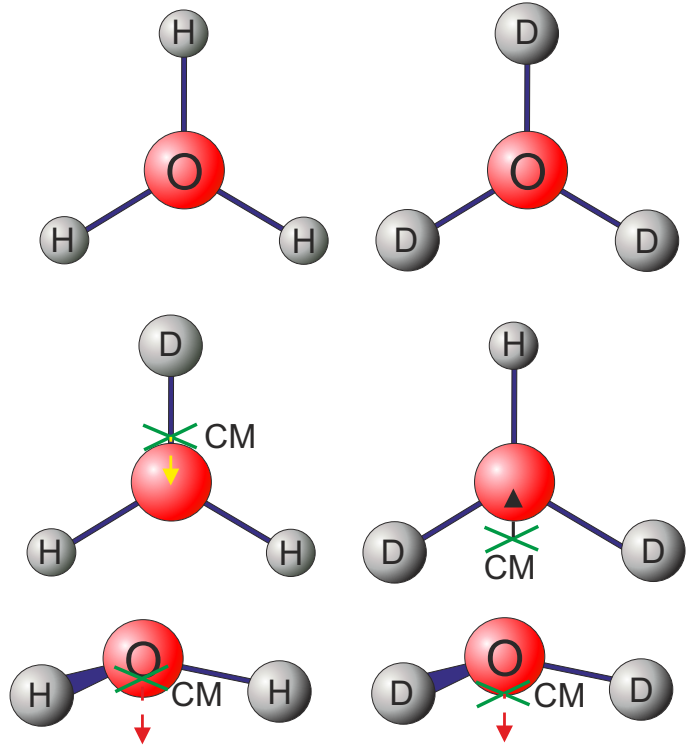}

  \caption{\label{fig:CM} Displacements of the center-of-mass (green crosses) upon
deuteration of \ohhh. Arrows indicate the dipole moment components.}

\end{figure}

As mentioned above,
the symmetry lowering from \Dh{3}(M) to \Cv{2}(M) gives rise to another important effect
illustrated in Fig.~\ref{fig:CM}. For \ohhh\ and \oddd\,   both the nuclear center-of-mass and
 the nuclear center-of-charge lie on the  $C_3$  symmetry axis for nuclear geometries with
\Cv{3} geometrical symmetry.\cite{98BuJexx.method}
We take the  $C_3$
axis  to be the $z$ axis
of the molecule-fixed axis system whose origin, by definition, is at the nuclear center-of-mass.
Consequently, at \Cv{3}-symmetry geometries the dipole moment
lies along the $z$ axis and its $x$- and $y$-components vanish.
The non-zero $z$-component  is responsible for the
parallel bands in the spectra of these species, including the rotation-inversion band \cite{09YuDrPe.H3O+} (the pure rotation band is forbidden by symmetry).
\citet{83BoRoRe.H3O+} estimated the corresponding transition dipole for the inversion $0^- \leftrightarrow 0^+$ band to be  1.47~D. Our \ai\ value is slightly higher, 1.80~D.
For
 \ohhd\ and \ohdd, the center-of-charge obviously is unchanged relative to
 \ohhh\ and \oddd\, but the center-of-mass is shifted, and this produces a non-vanishing
perpendicular dipole moment at \Cv{3}-symmetry geometries.
If we take the $x$ axis to be in the plane defined by the $C_3$ symmetry axis\cite{98BuJexx.method}
and the O--H bond for \ohdd, and
in the plane defined by the $C_3$ symmetry axis and the O--D bond for \ohhd, then
\ohdd\ and
 \ohhd\ acquire non-vanishing
 $x$
dipole moment components. Therefore the perpendicular transitions ($\Delta K_c=\pm 1$)
of \ohhd\ and \ohdd\ in the vibrational ground state are much stronger than
 the $\Delta K=\pm 1$ transitions
of \ohhh\ and \oddd\ which
are caused by intensity stealing from the vibrational spectrum.\cite{98BuJexx.method}
Besides, this $x$ component is larger for \ohdd\ than for \ohhd\ owing to
the greater displacement of the nuclear  center-of-mass. This is probably
why the \ohdd\ lifetimes are shorter on the average than        those of
\ohhd. The $z$ dipole moment component also changes with isotopic
substitutions, see Fig.~\ref{fig:CM}.

The longest-living states of \oddd\ (Table~\ref{tab:LF}) have lifetimes
about 27 times longer than those of \ohhh. Presumably,
this is mainly caused by the fact that
 \oddd\ has lower
 ro-vibrational term values than \ohhh.
Thus, \oddd\ has lower values of
 $\tilde{\nu}_{if}$ in Eq.~(\ref{e:A}) and this, in turn, causes
 lower Einstein-$A$ coefficients and  higher values of $\tau$ [Eqs.~(\ref{eq: tauval})
and (\ref{e:A})].

%The transitions to the H$_2$DO$^+$ and HD$_2$O$^+$ forms go with the $D_{3h} \to C_{2v}$ symmetry lowering, which leads to the significant reduction of lifetimes. A certain role here also plays dipole moment transformation, since for partly deuterated isotopologues the $x$-component of dipole moment vector becomes nonzero.

\begin{figure}%[!htbp]

  \includegraphics[width=9cm]{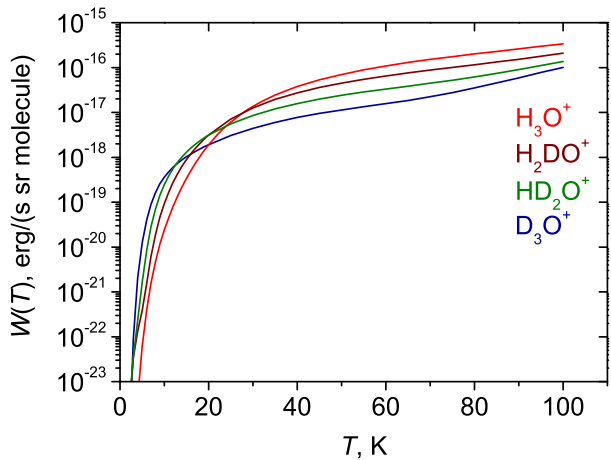}

  \caption{\label{fig:cooling} Calculated cooling functions of H$_3$O$^+$ and isotopologues.}

\end{figure}

The calculated radiative cooling functions $W(T)$
[Eq.~(\ref{e:cooling})] for H$_3$O$^+$ and
its deuterated isotopologues are shown
in Fig.~\ref{fig:cooling}. At  temperatures above 30~K
the cooling decreases with increasing numbers of deuterium
atoms in the molecule. This can be easily understood from Eqs.~(\ref{e:A}) and (\ref{e:cooling}):
$W(T)$ is proportional to $\tilde\nu_{if}^4$, and $\tilde{\nu}_{if}$ is
approximately inversely proportional to the mass of hydrogen for the rotational
states populated at the temperatures considered.
Therefore, at moderate and high temperatures the lighter isotopologues are better
coolers. However, Fig.~\ref{fig:cooling} shows that at lower temperatures their
roles change and the deuterated species become the better coolers.
This is because the term values of the
 lowest, infinite-lifetime (and therefore coldest) states      vary as
47.03~\cm\ $(2,2,E\p,0^+)$, 23.79~\cm\ $(2,2,E\p,0^+)$,
12.19~\cm\ $(1_{0,1}^{+},A_2)$, and
11.69~\cm\ $(1_{0,1}^+,A_2)$ for
 \ohhh, \oddd, \ohdd, and  \ohhd, respectively.
At very low temperatures, the molecules will tend to collect in the lowest accessible state,
and the higher the term value of this state, the more difficult it is to cool the isotopologue
in question. Because of this, for example,
  it is more difficult to cool \ohhh\ than \oddd\ at
temperatures below 30~K.

%\red{DOES NOT IT CONTRADICT THE COOLING FUNCTION?}

%\begin{figure}[!htbp]
%\begin{tabular}{cc}
%  \includegraphics[width=8cm]{h3op-2c.eps} & \includegraphics[width=8cm]{d3op-2c.eps} \\
%  \includegraphics[width=8cm]{h2dop-2c.eps}& \includegraphics[width=8cm]{hd2op-2c.eps}\\
%\end{tabular}
%  \caption{\label{fig:lifetime-2} Calculated lifetimes of the most long-living ro-vibrational states of H$_3$O$^+$ and isotopologues.  Energy states are labeled as $(J,K,\Gamma)$. \red{Vlad, please increase the font of axes title, units, labels (i.e. everything)  by a factor of 2}}
%\end{figure}

%-----------------------------------------------------------------------------
\section{Conclusions}

We have carried out
a theoretical study of the ro-vibrational states of the hydronium ion \ohhh\
and its deuterated isotopologues.
 \textit{Ab initio} potential energy and
electric dipole moment surfaces were used to calculate ro-vibrational energy
levels, corresponding wavefunctions and Einstein coefficients for the low-lying
ro-vibrational transitions of
 these ions.
We have analyzed stability of the ro-vibrational states and computed
the
 radiative lifetimes and cooling functions for temperatures below
200~K.

Taking into account only spontaneous emission as cause of decay of ro-vibrational
states (and neglecting collisions and stimulated
emission) we find the longest-lived hydronium state for
 \oddd: the population in the rotational state with $(J,K,\Gamma)$ $=$
$(5,5,E'')$  is trapped for 3816.0 years, which is relatively `hot'
% Another long-living state of \oddd\ is  $(2,2,E')$ (857.1 years), which is relatively `hot'
 (152~K), at least in the context of
 molecular cooling, for example in   storage rings. In this work we have identified a number of relatively hot ($E/k > 100$~K) meta-stable states with a lifetime longer than 10~s (typical timescales of ion storage experiments).  Such meta-stable states which will be populated and hamper the cooling of hydronium ions to a temperature of a few
 Kelvin. The molecule with
the shortest-lived meta-stable states is \ohdd\ with lifetimes of a few days.
 The timescale of interstellar collisions in diffuse clouds is longer (about a month), and thus some of these states undergo spontaneous emission.

Our calculations show that deuteration influences significantly
the hydronium lifetimes. This effect is mostly caused by the symmetry lowering
from
 \Dh{3}(M) to \Cv{2}(M)
and the ensuing perpendicular dipole moment component.
A  number of long-living meta-stable states are
identified, capable of                      population trapping.
Compared to the deuterated species,
the cooling
of the lightest isotopologue \ohhh\ is most efficient at
 higher temperatures
($T>30$~K). However, this changes at  very low
temperatures where the  \ohhh\ ions are trapped at relatively high energy.

The results obtained can be used to assess the cooling properties of the hydronium ion
in ion storage rings and elsewhere.
%Our calculations will be useful for estimating the cooling properties of hydronium ions, for %example, storage rings, ro-vibrational relaxation in the stored ion beam.

%-----------------------------------------------------------------------------
\section*{Acknowledgments}

This work was supported in part by ``The Tomsk State University Academic D.I. Mendeleev Fund Program'' grant No. 8.1.51.2015, and in part by the ERC under the Advanced Investigator Project 267219. SNY, JT, and PJ are grateful to  the COST action MOLIM (CM1405) for support. We thank Andreas Wolf for suggesting this work.

%\begin{thebibliography}{99}
%\baselineskip=24pt

%\input{paper16-1-refs.tex}

%\end{thebibliography}

%\bibliographystyle{elsarticle-num}
%\bibliographystyle{rsc}
%\bibliography{journals_phys,H3O+,SbH3,H3+,NH3,15NH3,linelists,methods,programs,PH3,SO3,SO3_,NH3p,CH2,CH3,H3+_,H2O_}

\providecommand*{\mcitethebibliography}{\thebibliography}
\csname @ifundefined\endcsname{endmcitethebibliography}
{\let\endmcitethebibliography\endthebibliography}{}

\end{document}